# APP STORE 2.0: From Crowd Information to Actionable Feedback in Mobile Ecosystems

María Gómez[1,2], Bram Adams[3], Walid Maalej[4], Martin Monperrus[2,1], Romain Rouvoy[2,1,5]

[1]Inria, France [2]University of Lille, France [3]Polytechnique Montreal, Canada [4]University of Hamburg, Germany [5]IUF, France

**Given the increasing competition in mobile app ecosystems, improving the experience of users has become a major goal for app vendors. This article introduces a visionary app store, called APP STORE 2.0, which exploits crowdsourced information about apps, devices and users to increase the overall quality of the delivered mobile apps. We sketch a blueprint architecture of the envisioned app stores and discuss different kinds of actionable feedback that app stores can generate using crowdsourced information.**

## I. DIVING INTO THE MOBILE ECOSYSTEM

Mobile devices, such as smartphones and tablets, are more and more infiltrating our daily activities. The widespread use of mobile devices has accelerated the development of mobile applications—broadly called *apps*. These apps are commonly distributed through platform-specific *app stores*, such as Google Play (Android), Apple Store (iOS), Windows Store (Windows Phone), etc. The stakes are huge: Google Play proposes 2.2 million apps while Apple Store surpasses 130 billion app downloads[1].

However, not all apps meet the quality requirements that users expect. For instance, app crashes and unresponsive apps severely disrupt user experience[2]. This is a major problem for app developers since previous studies have demonstrated that users who encounter issues (e.g. crashes) are likely to stop using the app. Even worse, negative reviews in early releases make it almost impossible to recover afterwards [1]. Thus, the major goal for app vendors is to detect and respond as fast as possible to quality issues, in particular crashes and unresponsive user interfaces.

The problem is not only a problem of having not invested enough resources in quality assurance. In the mobile app world, many bugs are independent of the care and effort put in ensuring high quality: there are many crashes and performance issues that come from the ecosystem itself. For instance, a platform API change may transform a good app into a crashing one. Another reason is the extremely high diversity of hardware devices, configuration settings and conditions of execution (*e.g.*, sensors, networks), which make it impossible to guarantee the proper functioning of a high-quality app in all situations.

Existing app stores offer limited support for helping app developers to detect and fix issues related to the ecosystem infrastructure itself (variety of APIs, OSes, hardware). Our key insight is that app stores should leverage the different types of crowds to which they have access—*crowd of apps*, *crowd of devices*, and *crowd of users*. In this article, we propose leveraging those crowds to engineer a new generation of app stores, coined APP STORE 2.0. The wisdom of those crowds can be combined, one augmenting the other, the sum being more powerful than each one in isolation.

TABLE I
TYPES AND VOLUMES OF CROWDS AND CROWDSOURCED INFORMATION IN APP STORES

| **Crowds** | | |
|---|---|---|
| Devices | 1,200+ | distinct device brands[3] |
|  | 24,000+ | distinct Android devices[3] |
| Apps | 2.2 million | distinct apps in Google Play[1] |
| Users | 3.79 billion | distinct mobile users[4] |
| **Crowdsourced information** | | |
| User reviews | 228+ million | user reviews in Google Play[4] |
| App logs | 60+/day/app | apps are run by 280 million users[5] |
| Device contexts | 23 | distinct Android OS API levels[2] |
|  | 1,411 | distinct requested permissions [2] |

APP STORE 2.0 considers three types of crowdsourced information (app reviews, app execution logs and app contexts). As reported in Table I, the volume, velocity and variety of this crowdsourced information meet the 3 V's of big data. APP STORE 2.0 uses this information to assist developers to deal with potential errors and threats that affect apps prior to publication and also when the apps are already in hands of end-users. The new app stores we envision are able to: 1) tell the developer about performance issues and regressions happening in the wild; 2) automatically synthesize tests that reproduce issues happening on specific devices or configurations only; 3) automatically infer the root causes of certain crashes such as those related to permissions; 4) warn app store administrators about bad apps that harm the app store reputation; 5) patch apps to prevent the occurrence of previously observed crashes.

To sum up, the crowd can contribute to the development of a new generation of app stores, 1) which would help developers to detect, diagnose and fix field bugs much faster; 2) which would help app store administrators to raise the average quality level of their stores; 3) which would enhance the overall user experience and satisfaction in the mobile digital world. In our recent research, we have prototyped and evaluated systems that provide those types of actionable feedback in the context of the Android ecosystem. Those ideas simply work in practice, without major changes in the infrastructure.

---

[1]http://www.statista.com/statistics/276623/number-of-appsavailable-in-leading-app-stores/

[2]Hewlett Packard. *Failing to meet mobile app user expectations: a mobile user survey.* Feb. 2015. http://bit.ly/1OOw5TB

[3]http://opensignal.com/reports/2015/08/android-fragmentation/

[4]http://wearesocial.com/uk/special-reports/digital-in-2016

[5]https://arc.applause.com/2015/12/16/applause-analytics-state-of-the-app-store-and-google-play-2015/

[6]http://flurrymobile.tumblr.com/post/124152019870/mobile-addicts-multiply-across-the-globe

## II. THE VISION OF APP STORE 2.0

Fig. 1 shows the APP STORE 2.0 blueprint architecture. The key idea of APP STORE 2.0 is to exploit the *wisdom of the crowd* to automatically increase the quality of the delivered mobile apps and provide actionable feedback to developers and users. The APP STORE 2.0 incorporates two types of quality-related actions:

1) **Pre-publication actions.** Upon submission of new apps, APP STORE 2.0 takes preventive actions: predictions of upcoming problems to avoid their actual occurrence in the field;
2) **Live actions.** When users install apps, if problems surface—such as crashes—APP STORE 2.0 immediately takes actions to assist the developer for fixing such issues.

### A. APP STORE 2.0 *Architectural Blueprint*

Current app stores, referred to as App Store 1.0 in this paper, have a Front Store service in charge of publication, browsing and delivery of apps. The typical workflow is shown as blue arrows in Figure 1. An app developer uploads an app to the store for distribution (steps 1, 2). Users download apps and execute apps on their devices (steps 3, 4). Additionally, users can write reviews about apps (step 5). The store provides developers with feedback regarding their apps, specifically user reviews and raw information regarding crashes and user experience issues, aka Application Not Responding errors (ANR).

The APP STORE 2.0 appends a new component, which we call Back Store service atop of the App Store 1.0 workflow. The Back Store component orchestrates a feedback loop which continuously supervises crowdsourced information (reviews, crash reports, execution logs) to detect and eventually fix defective apps. The Back Store consists of 5 modules: *Crowd Monitor*, *Risk Analyzer*, *Crash Analyzer*, *Performance Analyzer*, and *Patch Generator*. These modules work together to provide 4 types of actionable feedback: *risk reports*, *reproducible scenarios*, *performance reports*, and *app patches*.

The main APP STORE 2.0 workflow is shown as green arrows in Figure 1. Once an app is uploaded, the store runs a Risk Analysis to predict potential crashes before making the new app publicly available for users (step 2.1). If there is a risk of crash, the store sends a risk report to the store moderator and to the app developers.

After the app is downloaded and executed on the users' mobile devices, a Crash Analyzer component listens for crash occurrences in the wild (step 4.1). In the presence of crashes, the APP STORE 2.0 learns crash and context (*i.e.*, software and hardware configurations) patterns, which are turned into a reproducible scenario to help developers to quickly reproduce the observed errors.

When the developer uploads a new release, APP STORE 2.0 runs a Performance Analysis to ensure that the new release does not perform worse than the previous release (step 2.2). A detailed performance report is provided to the developer, who can update the app to fix the performance defects prior to its publication.

Meanwhile, as the fix and release process can be long [3], the store generates hot patches (step 4.2) and updates apps to prevent other users to suffer from the crashes again and again (step 4.3). In other words, the store keeps monitoring the information crowdsourced from devices and user feedback as an oracle for the autonomous improvement process.

Three key audiences can benefit from those actionable feedback:

- *App stores*. Currently there is a huge range of app stores available, all competing to attract customers (developers and users). App stores can implement the presented APP STORE 2.0 approach to enhance their services, improving the quality of apps and consequently the store's reputation.
- *App developers*. Developers want to deliver high quality apps to survive the market competition. By using this approach, they can increase the quality of their apps, thus contributing to improve their users' satisfaction and loyalty.
- *App users*. Users want high-quality apps that ensure a high user experience.

---

**Related Work**

**App Store Analysis**
App Stores emerged in 2008 with the launch of the Apple App Store. Since then, a vast variety of studies have investigated those rich repositories. Previous research have revealed the diversity of information provided by user feedback in app stores [4]. Developers can benefit from this knowledge to improve their apps. William et al.'s present an exhaustive survey which covers all published literature on app store analysis up-to-date [5]. These studies span over *review analysis*, *API usage*, *feature analysis*, *release engineering*, *security*, *prediction*, and *store ecosystem*.

**Involving Users in Software Lifecycle**
Maalej and Pagano proposed the first framework to involve users and user communities in software engineering processes [6]. Recent research seek to employ data-driven techniques (based on user feedback) to support requirement and software engineering tasks [7].

---

## III. REALIZING THE APP STORE 2.0

We now transform the vision of APP STORE 2.0 into a concrete implementation for Android. In the following sections, we describe the prototype tools we have implemented to support each of the 4 modules which compose the APP STORE 2.0: *Risk Analyzer*, *Crash Analyzer*, *Performance Analyzer*, and *Patch Generator*. Fig. 2 summarizes the supporting tools. These modules exploit 3 types of crowd sources (*user reviews*, *crash reports*, *execution logs*) to generate 4 different types of actionable feedback that can assist developers to improve the quality of their mobile apps. We now present those modules.

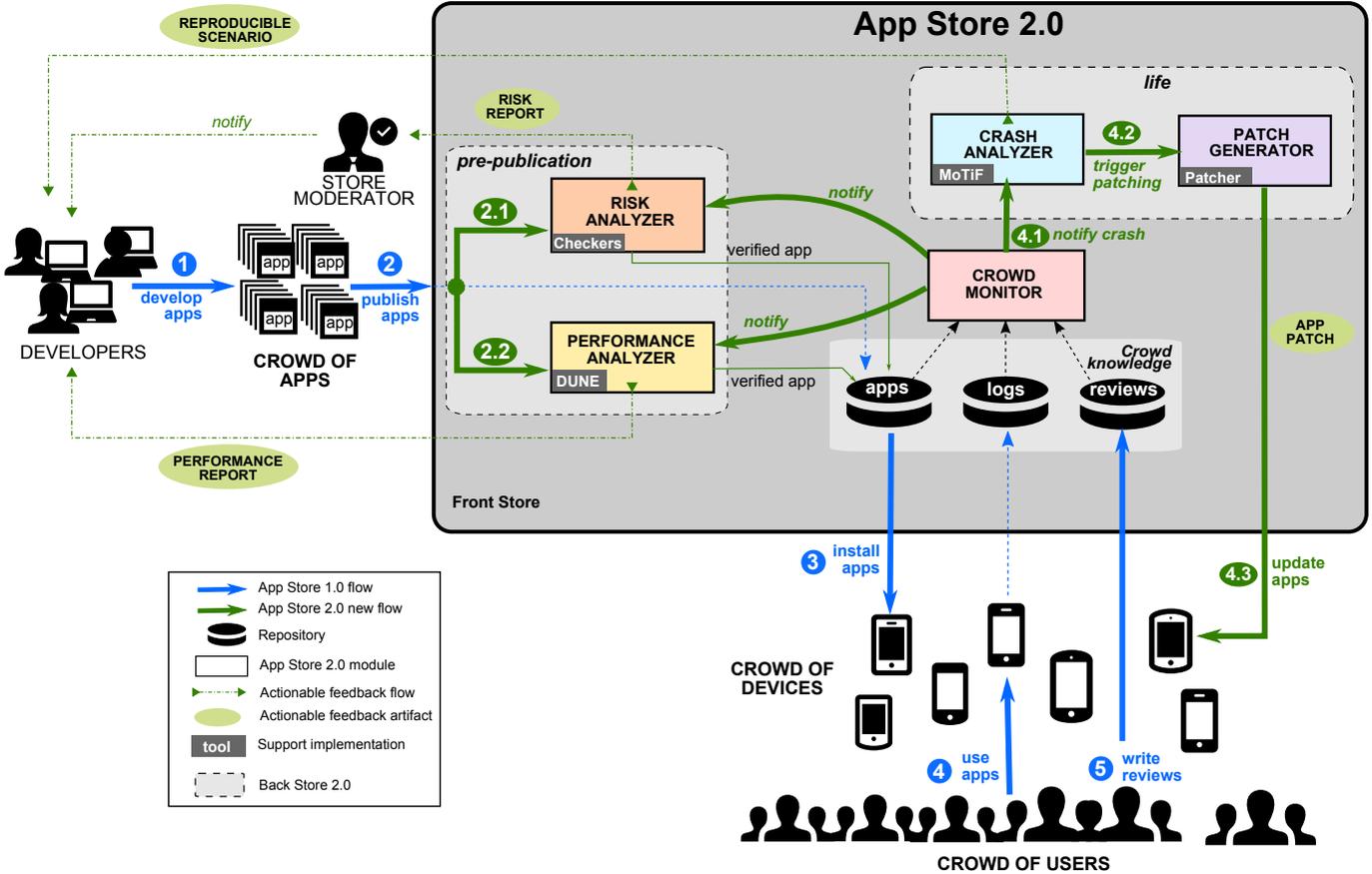

Fig. 1. The APP STORE 2.0 reference architecture

### A. Reporting Risky Apps a priori

The APP STORE 2.0 incorporates *crowd-based checkers* to rank the risk of a crash in newly submitted apps [2]. The Risk Analyzer component builds *checkers* based on observations of *user reviews* of existing apps executed by the crowd, and the associated permissions requested by the apps. The process to build the *checkers* comprises three steps (cf. Fig. 2 III-A):

*1) Analyzing User Feedback*

The Crowd Monitor continuously supervises user reviews published in the store to identify apps that tend to crash in the hands of end-users. It works as follows. First, the system extracts topics discussed in the corpus of reviews using *Topic Modeling* (in particular the *Latent Dirichlet Allocation*—LDA—technique). Second, the system classifies as *crash-related reviews* those that are mainly composed of topics related to crashes and bugs. Lastly, it flags as *crash-prone* the apps whose ratio of error-related reviews reaches a predefined threshold.

In the dataset of $46,644$ apps and their $1,402,717$ user reviews that we collected from the Google Play Store, our system enabled the identification of $10,658$ crash-prone apps (cf. [2] for implementation details).

*2) Analyzing App Permissions*

Android apps need to explicitly request permissions to use APIs that give access to protected system resources (*e.g.*, data, privileged operations) and third-party libraries. Our insight is that some crashes are correlated with permissions. For instance, the use of buggy or obsolete APIs can lead to crashes on new devices. In other words, we use the permissions requested by apps as a proxy for buggy functionality. Once the store identifies a cluster of crash-prone apps, it notifies the Risk Analyzer component which searches for recurring bad permission patterns that correlate with the crashes. The Risk Analyzer component creates a predictive machine-learning model (a *J48* Decision Tree) to predict if an app will likely crash based on the set of requested permissions.

*3) Generating Risk Reports*

The resulting crash prediction model constitutes the basis of the *crowd-based checkers* embedded in the store. App store moderators can activate these checkers to score the risk of crashes of newly submitted apps. As all checkers, app store checkers may suffer from false positives. A false positive is when an app is flagged as being potentially buggy while it actually works fine. If an app store moderator enables checkers with too many false positives, it would be a deal-breaker for app developers. On the other hand, if all checkers are disabled, the store risks hosting buggy apps, which would degrade the store reputation and app popularity. For this reason, if there is a risk of crash a *risk report* is sent to the store moderator, who then decides to publish the app or to notify the developer. Note that, as the app ecosystem is continuously evolving, app stores can run our approach regularly (say weekly) to update existing checkers, discover new ones, and discard outdated ones.

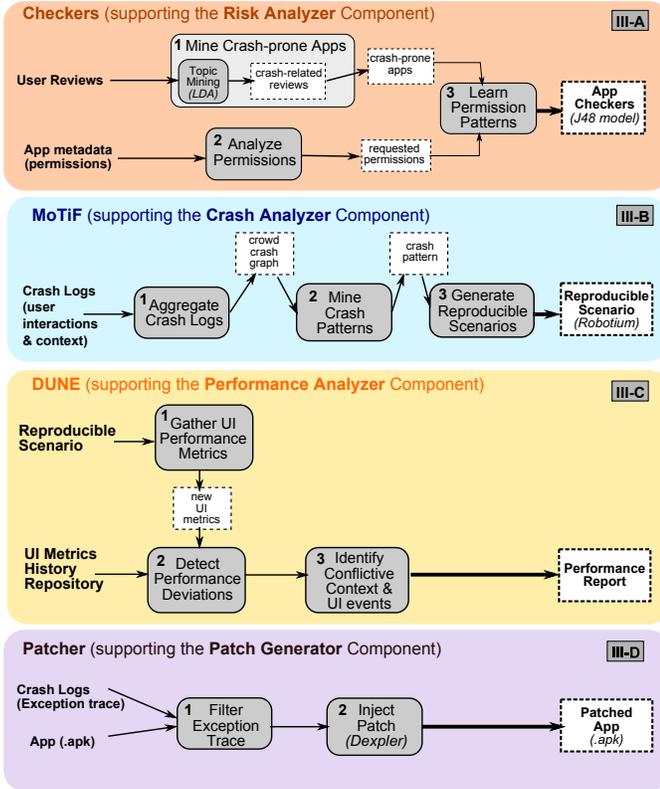

Fig. 2. Implementation of the APP STORE 2.0 components

Using our Google Play Store dataset, we have built a family of checkers from user reviews and permission requests (cf. [2] for implementation details). These checkers successfully predicted crashes caused after the update of *Google Play services 4.3* (March 2014).

### B. Reproducing Crash Scenarios a posteriori

Software developers know that faithfully reproducing crashes experienced by users in the wild is a major challenge. Crash reproduction is even harder in mobile environments due to the high heterogeneity of hardware, mobile platform releases, and execution contexts. Once an app is published and downloaded on the users' devices, the APP STORE 2.0 receives crash reports coming from the execution of apps. The Crash Analyzer component exploits crashes to isolate crash conditions and to reproduce crashes in an automatic and effective manner. It includes 3 steps (cf. Fig. 2 III-B):

*1) Monitoring App Executions*

The APP STORE 2.0 implements a two-level monitoring strategy. Initially, the Crash Analyzer component listens for crashes happening during the execution of apps. When a certain ratio of users suffer from crashes, the store flags the app as *defective*. Then, the system activates a lightweight monitoring mechanism to gather additional *execution logs* of user interactions. An *execution log* contains a sequence of user interactions (such as clicks) and operating contexts—*i.e.*, *static context* (such as device model, manufacture and SDK version) and *dynamic context* (such as state of sensors, battery, network) observed during the execution.

In the APP STORE 2.0, the monitoring is distributed on the crowd of devices running a defective app. To avoid any accidental user disturbance, the monitoring is periodically redistributed among users in the crowd and only one app is monitored on each device. Furthermore, app developers, when submitting apps to the store, can set the threshold of users subject to advanced logging. This is in contrast to previous research which proposes to monitor the interactions of all users [8].

*2) Mining Crash Patterns*

Afterwards, the Crash Analyzer component aggregates the crowdsourced execution logs of user interactions and contexts into a graph. This graph provides an aggregated view of interactions of a multitude of users of a defective app. The graph is then used to identify patterns of interactions and contexts that appear frequently among crashes. The Crash Analyzer component implements different data-mining techniques (*path analysis*, *sequential patterns*, and *set operations*) to effectively infer the minimal sequence of interactions that recreates a crash, as well as the context under which the crash occurs.

*3) Generating Reproducible Scenarios*

Finally, the minimal sequence of interactions is translated into a *reproducible scenario* to automatically recreate the crash faced by users. Thi *reproducible scenario* is implemented as a *Robotium* black-box UI test[7].

Before providing the *reproducible scenario* to developers, the store replays the scenarios on a sample of devices in the crowd to assess whether or not 1) the scenario truly reproduces the observed crashes, and 2) the scenario generalizes to other contexts or devices (*e.g.*, not all devices suffer from the same bugs). During the execution of the scenarios for validation, the store additionally logs UI performance metrics (*i.e.*, frame rate) to populate a repository of *historical executions* for different context configurations. Note that, to avoid any user disturbance, the APP STORE 2.0 executes the scenarios for validation only during periods of phone inactivity (*e.g.*, during the night, and when the device is charging).

We have implemented a prototype tool, MOTIF, to support this module. Using this tool, we performed an experiment with 10 users, showing that APP STORE 2.0 is able to reproduce 4 out of 5 real crashes in Android apps (cf. [9] for details).

---

[7]http://www.robotium.org

## C. Reporting on Performance Degradations

One major non-functional requirement of mobile apps is to guarantee smooth user-interface (UI) interactions. UI smoothness defects are known in the Android developers community as *janks*. The main research challenge of automatically identifying janks on mobile devices is that the UI performance of an app highly varies depending on its context. For example, an app can perform well on a set of devices, but it may exhibit janks in a different environment consisting of, amongst others, a different device model or OS version.

Once an app developer has used the *reproducible scenario* generated by APP STORE 2.0 to fix the crashes and to upload a new release of her app to the store, the Performance Analyzer component (cf. Fig. 2 III-C) receives as input the *reproducible scenario* and the *execution history repository* (filled by the Crash Analyzer component). Upon new app release, the system repeats the *reproducible scenario* with the new app release on different devices of the crowd, while collecting UI performance metrics. The goal is to assess whether the new app release fulfills the expected performance goals—*i.e.*, to prevent performance degradations and to improve performance on low-end devices. Then, it compares the newly collected performance metrics against the previous metrics available in the *execution history repository* to automatically flag performance deviations.

The process to detect performance deviations is as follows. To compare an execution with the historical executions, the Performance Analyzer component calculates a *context similarity* and applies a statistical technique (in particular *Interquartile Range*) to flag a performance regression when the distance between the performance metrics of the new and old executions in similar contexts is larger than a given threshold. If the new release is flagged as an outlier, the system identifies the device configurations and specific UI events that trigger the performance deviations. A detailed report is sent to the developer who can fix the app before publication.

To demonstrate the feasibility of this module, we have implemented a prototype called DUNE. Using DUNE we were able to identify real performance degradations (in the *K9 Mail*[8] and *Space Blaster*[9] Android apps) reported by users on Android 5.0 devices (cf. [10] for additional information).

## D. Patching Defective Apps in the Wild

While the developer is working on fixing the app, the Patch Generator component generates temporary patches to prevent recurrences of the same crash for different users. The goal of this component is to synthesize candidate patches using automated repair approaches [11]. For each synthesized patch, the Patch Generator component creates a new release of the defective app that includes the patch.

When the store receives a download request for an app that has been previously flagged as defective, the store delivers an alternative patched release of the app. Afterwards, the store keeps monitoring crowdsourced information from devices and user feedback that run those automatically patched apps to assess the effectiveness of the generated patches. If a patch generation technique fails (*e.g.*, the patched app still crashes), the store searches for a better alternative patch. The APP STORE 2.0 continuously monitors crowdsourced information to improve the patching process.

We have implemented a *Patcher* which provides a patching strategy for exceptions (cf. Fig. 2 III-D). The *Patcher* takes as input the *exception trace* thrown by an app crash and extracts *suspicious methods* that appear in the exception trace. A patch wraps the code defined inside the suspicious methods with a `try/catch block` to capture the runtime exceptions that are not handled by the methods. The system creates different patched versions of the defective app, where each patch wraps a different suspicious method. They are all then tested in the wild. To inject the patches, our *Patcher* instruments the bytecode of the Android apps using *Dexpler* [12]. Using this technique we could prevent a crash in an Android app reported as buggy by users (cf. [13] for additional information).

Further details about the implementations are available online: http://app-store.apisense.io.

> **Techniques**
>
> - **Topic Modeling:** Technique to extract topics from a corpus of unlabeled text. A topic is a list of words that occur frequently together along the texts. **LDA** *(Latent Dirichlet Allocation)* is a specific topic modeling technique. APP STORE 2.0 uses topic modeling on user reviews.
> - **J48 Decision Tree:** A machine learning algorithm which generates decision trees for classification. APP STORE 2.0 uses decision trees to predict crashes related to permissions.
> - **Path Analysis:** In graph theory, path analysis techniques find a path between two nodes according to a criteria, e.g. finding shortest path. APP STORE 2.0 uses path analysis to infer a minimal sequence of interactions that produces a crash.
> - **Frequent Itemset Mining:** Data mining technique to find frequent patterns in data.
>   **Sequential Pattern Mining** is a specific frequent itemset technique to find data that are frequently in sequence. APP STORE 2.0 uses sequential patterns mining to identify conditions correlated with crashes.
> - **Interquartile Range:** Statistical technique to filter outliers and extreme values based on interquartile ranges. APP STORE 2.0 computes interquartile ranges to identify user-interface performance regressions.

## IV. PRACTICAL IMPLICATIONS AND LIMITATIONS OF APP STORE 2.0

The barrier to adopting APP STORE 2.0 is low for developers because they are already used to include crash-reporting libraries in their apps to obtain information about failures. The realization of the APP STORE 2.0 only requires the inclusion of a more sophisticated library, which would include user interaction logging, performance analysis, ...).

---

[8] https://play.google.com/store/apps/details?id=com.fsck.k9
[9] https://play.google.com/store/apps/details?id=com.iraqdev.spece

Nevertheless, the APP STORE 2.0 raises two critical concerns: user privacy and security. Regarding privacy, the APP STORE 2.0 logs a large number of information about users and their activity. Consequently, the APP STORE 2.0 needs to establish privacy policies to reach agreements between developers, users, and stores. End-users must have a way to configure their privacy preferences to grant access to the types of data that can be collected and the phases of the process where they volunteer to participate. Furthermore, the APP STORE 2.0 needs to enforce security protocols to ensure that any malicious app can bypass security measures and take control of the devices.

Finally, the APP STORE 2.0 creates value based on the end-users. While the end-users indirectly benefit from APP STORE 2.0 with better apps, we also imagine more direct ways to involve users in the automated quality feedback loop. In this direction, we imagine that the APP STORE 2.0 would provide incentive mechanisms [14] to encourage users to participate in the process of collecting execution data, running performance measurement scenarios, executing automatically generated patched versions. For instance, users can be rewarded with early access to new updates, with free access to paid functionalities, etc.

## V. CONCLUSION & PERSPECTIVES

App stores have access to an enormous amount of crowdsourced information. Our vision of APP STORE 2.0 is to exploit thus crowdsourced information to automatically improve the user experience in the mobile digital world. APP STORE 2.0 provides an application ecosystem with the following benefits. First, it improves the speed to detect problems in the wild (crashes, performance problems). Second, it prevents bad apps to reach the mass of users. Third, it automates the bug diagnosis and fixing process, thus reducing human intervention to maintain mobile apps.

Beyond the specific vision and realization of APP STORE 2.0, we refer the reader to Nagappan et al. [15] for the community-identified future directions in software engineering for mobile apps.